\documentclass[a4paper, amsfonts, amssymb, amsmath, reprint, showkeys, nofootinbib, twoside]{revtex4-1}
\usepackage[english]{babel}
\usepackage[utf8]{inputenc}
\usepackage[colorinlistoftodos, color=green!40, prependcaption]{todonotes}
\usepackage{amsthm}
\usepackage{mathtools}
\usepackage{physics}
\usepackage{xcolor}
\usepackage{graphicx}
\usepackage[left=23mm,right=13mm,top=35mm,columnsep=15pt]{geometry} 
\usepackage{adjustbox}
\usepackage{placeins}
\usepackage[T1]{fontenc}
\usepackage{lipsum}
\usepackage{csquotes}
\usepackage{booktabs}
\usepackage{verbatim}

\usepackage{tabu}
\usepackage{xcolor}
\usepackage{tabularx}
\usepackage{booktabs}
 
\usepackage[pdftex, pdftitle={Article}, pdfauthor={Author}]{hyperref} 
\newcommand{\BFO}[0]{BiFeO$_3$ }
\newcommand{\BFOo}[0]{BiFeO$_3$}

\begin{document}
\title{Prediction of new low-energy phases of \BFO with large unit cell and complex tilts beyond Glazer notation}

\author{Bastien F. Grosso}
    \email[Correspondence email address: ]{bastien.grosso@mat.ethz.ch}
    \author{Nicola A. Spaldin}

    \affiliation{Materials Theory, ETH Zürich, Wolfgang-Pauli-Strasse 27, 8093 Zürich, Switzerland}
    
\date{\today} 


\begin{abstract}
Bismuth ferrite is one of the most widely studied multiferroic materials because of its large ferroelectric polarisation coexisting with magnetic order at room temperature. Using density functional theory (DFT), we identify several previously unknown polar and non-polar structures within the low-energy phase space of perovskite-structure bismuth ferrite, \BFOo.  Of particular interest is a series of non-centrosymmetric structures with  polarisation along one lattice vector, combined with anti-polar distortions, reminiscent of ferroelectric domains, along a perpendicular direction. We discuss possible routes to stabilising the new phases using biaxial heteroepitaxial strain or interfacial electrostatic control in heterostructures.
\end{abstract}

\maketitle

\section{Introduction} \label{sec:introduction}
Bismuth ferrite, \BFOo, is one of the few materials that combines magnetic order and ferroelectricity in the same phase at room temperature, making it one the most well-studied multiferroics. The structural ground state of \BFO is a distorted $R3c$-symmetry perovskite, with anti-ferrodistortive rotations of the oxygen octahedra around the pseudo-cubic [111] axis, combined with a large polarisation ($\sim$ 90 $\mu$C/cm$^2$) along the [111] direction caused by the $6s^2$ lone pairs of electrons on the Bi$^{3+}$ ions \cite{Wang/Neaton/Zheng:2003}. The strong superexchange between the Fe$^{3+}$ $3d^5$ electrons gives robust antiferromagnetism.

While the $R3c$ structural ground state is both theoretically \cite{Wang/Neaton/Zheng:2003} and experimentally \cite{Kubel/Schmid:1990} well established, the low-energy structural phase space of \BFO is known to be very rich. Under compressive strain imposed by coherent heteroepitaxy, polar monoclinic, tetragonal-like and tetragonal structures can be stabilized \cite{Bea/Dupe/Fusil:2009,Zeches/Rossell/Zhang:2009,Christen/Nam/Kim:2011,Hatt/Spaldin/Ederer:2010}, whereas under tensile strain or hydrostatic pressure it adopts a polar orthorhombic structure \cite{Yang/Ren/Stengel:2012,Yang/He/Suresha:2012,Kozlenko/Belik/Belushkin:2011}. More recently, heteroepitaxy has been exploited to form antipolar phases of \BFO in superlattices with La$_{0.7}$Sr$_{0.3}$MnO$_3$  \cite{Dong/Peters/Rusu:2020} and La$_{0.4}$Bi$_{0.6}$FeO$_3$ \cite{Mundy/Heikes/Grosso:2018}, where the stability of the antipolar phase was attributed to magnetic coupling and control of the electrostatic boundary conditions respectively. Furthermore, a systematic density functional theory (DFT) study revealed a large number of low-energy metastable structures within unit cells up to a size of 40 atoms \cite{Dieguez/Gonzalez-Vazquez/Wojdel:2011}.

Motivated in particular by the recent observations of ultra-large unit cell phases in thin-film heterostructures, we present a detailed computational study of the low-energy structural phase space of \BFOo, with a focus on large-unit-cell structures. Our approach is to start from the previously identified phases of Ref.~\onlinecite{Dieguez/Gonzalez-Vazquez/Wojdel:2011} and search for unstable phonon modes that indicate structural instabilities. We then increase the unit cell size to accommodate the corresponding energy lowering structural distortion. With this approach, we find that the apparent metastability of many of the previously identified structures, for example $Pbam$ and $Pmc2_1$, was an artifact of the 40-atom unit cell size, and in fact their energy is lowered by large period structural distortions. Finally, we demonstrate how these new phases can be stabilized relative to the bulk $R3c$ ground state through careful engineering of the electrostatic and strain boundary conditions, facilitating the rational design of \BFOo-based heterostructures with new functionalities.

The remainder of this paper is organized as follows: In section \ref{sec:methods}, we present the technical details of the simulations and introduce a generalisation to the standard Glazer notation \cite{Glazer:1972}, that allows the description of more complex oxygen octahedral tilt patterns. In section \ref{sec:phases_distortions}, we present the new phases that we identify in this work and describe their structures in terms of their main distortion modes. In section \ref{sec:nanometers_uc}, we consider a series of structures obtained by increasing the unit cell length along one direction and compare this series of structures with the formation of domains. In section \ref{sec:stability}, we study the possible stabilisation of the new phases using strain or control of the electrostatic boundary conditions. Finally, in section \ref{sec:conclusion}, we summarize our study.

\section{Methods} \label{sec:methods}
\subsection{Computational details}
The calculations were performed using density-functional theory (DFT) \cite{Kohn/Sham:1965} with the projector augmented wave (PAW) method \cite{Blochl:1994} as implemented in the Vienna ab initio simulation package (VASP 5.4.4) \cite{Kresse/Furthmuller:1996}. We used a 12x12x12 k-point $\Gamma$-centered mesh to sample the Brillouin zone corresponding to a 5-atom unit cell, and chose and energy cutoff of 850 eV for the plane-wave basis. The following valence electron configurations where used \footnote{The Bi, Fe and O PAWs are dated respectively from April 8 2002, September 6 2000 and April 8 2002.}: $5d^{10}6s^26p^3$ for bismuth, $3p^63d^74s^1$ for iron and $2s^22p^4$ for oxygen. We used the PBEsol$+U$ functional form of the generalized gradient approximation \cite{Perdew/Burke/Ernzerhof:1996}, with a commonly used value of $U_{\text{eff}}$ = 4 eV for the Fe $3d$ orbitals, according to Dudarev's approach\cite{Dudarev/Botton/Savrasov:1998}.

\subsection{Phonons and symmetry}
In order to identify new metastable phases, we start from the following experimentally stabilized, or computationally predicted polymorphs of \BFOo : $Pmc2_1$ (20 atoms/unit cell (u.c.)), $Pbam$ (40 atoms/u.c.) and $Pnma$ (20 atoms/u.c.) \cite{Yang/Ren/Stengel:2012,Kozlenko/Belik/Belushkin:2011,Jain/PingOng/Hautier:2013}) and explore the phonon instabilities along various reciprocal directions. We use the frozen-phonon method, as implemented in the \emph{PHONOPY} package \cite{Togo/Tanaka:2015} to look for imaginary phonon frequencies at selected symmetry points corresponding to supercells up to 160 atoms per unit cell. In practice, for a given starting phase, we construct all different possible supercells within our limit on the number of atoms and compute the phonon frequencies to look for instabilities. Once an instability is found we freeze the distortions given by the eigenvector corresponding to that imaginary frequency and fully relax the structure; if this procedure identifies a new (meta)stable phase, we repeat the procedure on that phase iteratively. We restrict our subsequent analysis to metastable phases with an energy up to 100 meV/f.u. above the $R3c$ ground state. We further test the stability of the relaxed structures (up to 80 atoms per unit cell) by computing the phonons at the zone center and specify the cases where an instability remains. We analyse each fully relaxed structure to determine the combination of phonon modes that contribute to its structural distortion from the reference cubic perovskite structure, using the \emph{ISODISTORT} \cite{Stokes/Hatch/Campbell,Campbell/Stokes/Tanner:2006}, \emph{AMPLIMODES} \cite{Orobengoa/Capillas/Aroyo:2009,Perez-Mato/Orobengoa/Aroyo:2010} and Pymatgen \cite{PingOng/DavidsonRichards/Jain:2013} packages. 

\subsection{Born effective charges and Polarisation}\label{sec:born}
Since Berry phase calculations of the polarisation are prohibitively expensive for very large unit cells, we compute polarisations by summing the displacement of the ions from a high-symmetry non-polar parent structure multiplied by their Born effective charges. We use an averaged Born effective charge per atom type obtained from averaging over the tensor components for 15 structures, including all phases presented in this work with 80-atom unit cells or less and other higher energy phases, computed in this work but not reported, to obtain one value per atom species. This procedure results in the following values: 4.86 [e] for Bi, 3.99 [e] for Fe and -2.95 [e] for O in units of the electronic charge magnitude, consistent with the literature. \cite{Neaton/Ederer/Waghmare:2005} We observe that the Born effective charges decrease substantially with structural distortions to more stable structures as the band gap energy increases.

\subsection{Extended Glazer notation for complex tilting} \label{sec:glazer}
Glazer introduced a convenient and widely used method for describing octahedral tilting in perovskites \cite{Glazer:1972}, which consists of a letter ($a$, $b$ or $c$) indicating the amplitude of tilting along each of the cartesian directions and a superscript $+$, $-$ or 0, to indicate whether the relative tilts between consecutive octahedra are the same, opposite or zero about the respective axis. This notation is restricted to tilt patterns with wave-vectors $k = 0$ or $\frac{\pi}{a}$, where $a$ is the lattice vector along the considered direction. More complicated tiltings, not captured by the current Glazer notation have been reported \cite{Peel/Thompson/Daoud-Aladine:2012,Xu/Wang/Iniguez:2014} and are prevalent in this work. We propose the following extension to Glazer notation suitable for generic tilting: we retain the $a$, $b$ and $c$ notation referring to each lattice vector, but we replace the $+$, $-$ or 0 superscripts by a series of Greek letters $(\alpha, \beta, \gamma, ...)$ indicating the magnitude of rotation of each octahedron around the pseudo-cubic direction considered. An over-line then indicates clockwise rotations, no overline indicates anti-clockwise and a $\emptyset$ no rotation. The number of superscripts indicates the number of octahedra in the periodic repeat unit and letters are repeated in the series if the rotation amplitudes of different octahedra are equal. If the exact same sequence of amplitude of the tilts is observed along different directions, the $a$ or $b$ are repeated along with the series of superscripts. Due to the periodicity of these series in a crystal, there are multiple ways to write the same sequence; we make the choice to always start the sequence by a clockwise rotation, if there is one. 

For simple tilt patterns the connection between the traditional Glazer notation and the extended Glazer notation can be easily made, for example: 
\vspace{0.1cm}
\begin{center}
\begin{tabular}{l l c l}
$Pnma$ & \hspace{0.05cm} $a^-a^-c^+$ & $\Longleftrightarrow$ & $a^{\bar{\alpha} \alpha}$ $a^{\bar{\alpha}\alpha}$ $c^{\bar{\alpha} \bar{\alpha}}$ \\
$R3c$ & \hspace{0.05cm} $a^-a^-a^-$ & $\Longleftrightarrow$ & $a^{\bar{\alpha}\alpha}$ $a^{\bar{\alpha}\alpha}$ $a^{\bar{\alpha}\alpha}$ \\
$Immm$ & \hspace{0.05cm} $a^0b^+c^+$ & $\Longleftrightarrow$ & $a^{\emptyset}$ $b^{\bar{\alpha}\bar{\alpha}}$ $c^{\bar{\alpha}\bar{\alpha}}$ \\
$Pm\bar{3}m$ & \hspace{0.05cm} $a^0a^0a^0$ & $\Longleftrightarrow$ & $a^{\emptyset}$ $a^{\emptyset}$ $a^{\emptyset}$ 
\end{tabular}
\end{center}
\vspace{0.1cm}

\noindent In this case, the extended notation offers no advantage over the traditional one. For more complicated tilts, with different amplitudes or complex sequences, it provides a convenient notation. For example, a tilting characterized along two directions by a wave-vector of $\frac{\pi}{4a}$, with two octahedra tilted in one direction (with same amplitude), the next pair tilted in the opposite direction with a different tilting angle from the first pair (and with different amplitudes among the pair) and a simple opposite tilting (with same amplitude) along the third direction is written $a^{\bar{\alpha}\bar{\alpha}\beta\gamma}a^{\bar{\alpha}\bar{\alpha}\beta\gamma}c^{\bar{\alpha}\alpha}$ in this notation. Notice that one could use different Greek letters along different directions to signal different (or similar) rotational amplitudes if this extra information were of interest, but we make the choice not to distinguish the amplitudes along different directions. 

We report the tilt patterns for the phases that were identified in this study, using this extended notation in Table \ref{tab_abc_E_P}.
\section{Results} \label{sec:results}

\subsection{Phases and distortions} \label{sec:phases_distortions}
We now follow the procedure described in section \ref{sec:methods} to identify various new polymorphs of \BFOo. We extend previous studies \cite{Dieguez/Iniguez:2011}, by exploring unit cells up to 160 atoms in size, and of variable shape; this allows us to incorporate phonon instabilities at lower symmetry q-points in the Brillouin zone. 
\begin{center}\
\begin{figure}[h]
\includegraphics[width=\columnwidth]{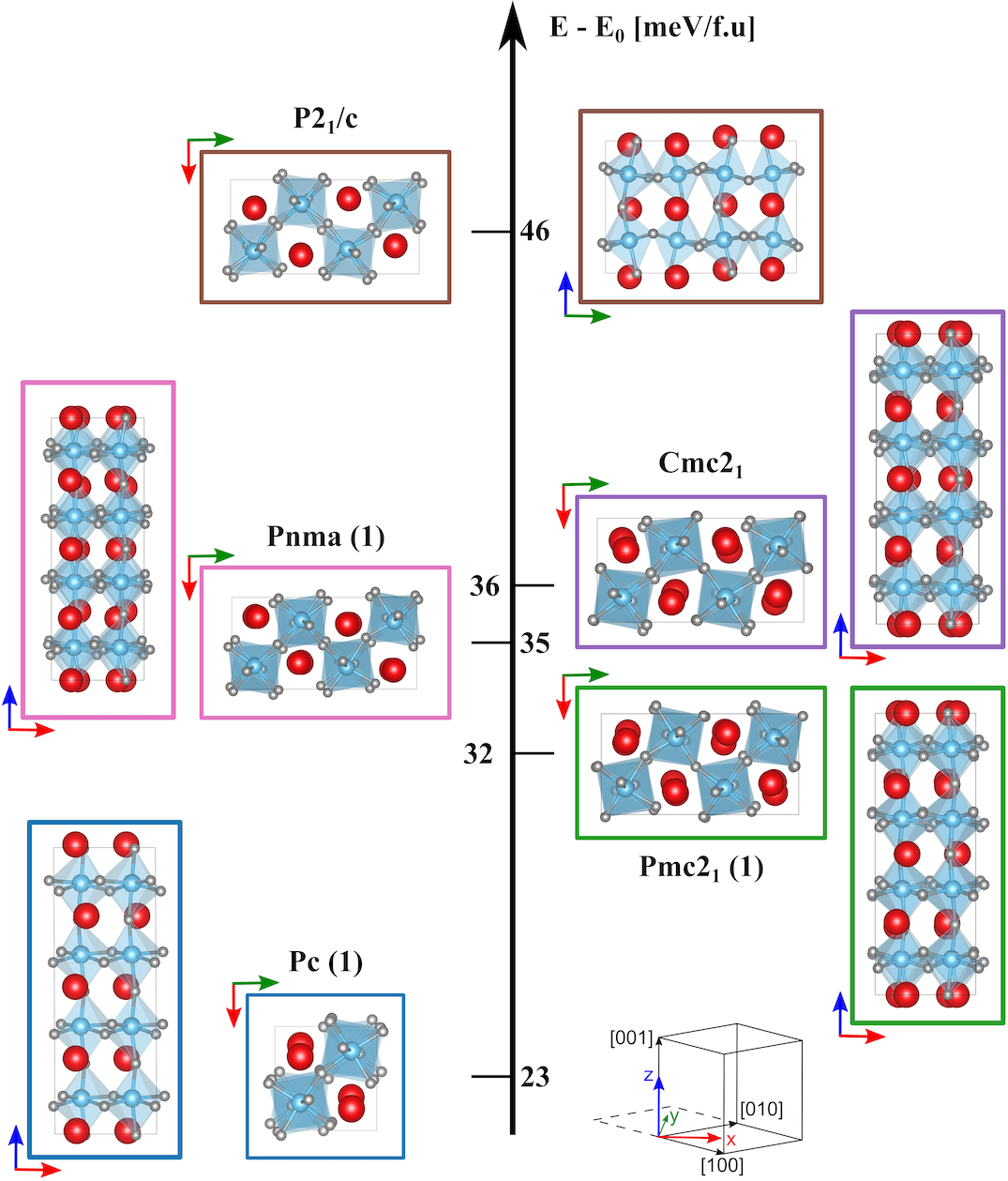}
\caption{\textbf{Crystal structures of new low-energy phases identified by full relaxation using DFT - } The vertical axis indicates the energy relative to the ground state. We orientate the structures such that the z axis stands along the usual [001] growth orientation. The Bi atoms are represented in red, the Fe atoms in blue and the O atoms in grey.}\
\label{fig_structures}
\end{figure}\
\end{center}\
In Table \ref{tab_abc_E_P} we show the list of newly identified phases that have an energy difference of less than 100 meV/f.u. from the $R3c$ ground state, and report the lattice parameters, energy relative to the ground state, band gap and polarisation along each lattice vector for each case. In Table \ref{tab_modes}, we decompose each phase into its main irreducible distortion modes and report its tilting pattern in terms of the traditional Glazer notation when possible and in terms of the extended notation introduced earlier (all cases). For reference we include the ground state ($R3c$) and the lowest energy non-polar ($Pnma$) phase in both tables. In Figure \ref{fig_structures} we present the unit cells of the five lowest energy new phases, and in Figure \ref{fig_distortions}, the dominant distortion modes that make up the structures, with the presiding distortions highlighted by arrows.

\vspace{0.5cm}
Before starting the analysis of the new phases identified in the current study, we summarize briefly the well-established reference phases for \BFOo. The $R3c$ ground state is reached from the ideal cubic perovskite structure via a polar displacement along the $[111]_{pc}$ direction ($\Gamma^{-}_{4}$ mode) and a tilting of the octahedra around the same direction ($R_5^-$ mode) resulting in an $a^-a^-a^-$ tilt pattern. At high temperature, \BFO adopts the GdFeO$_3$-like $Pnma$ phase,\cite{Geller:1956} characterised by antipolar displacements of the Bi atoms accompanied by anti-phase in-plane rotations and in-phase out-of-plane tiltings of the octahedra, resulting in a $a^-a^-c^+$ tilt pattern. This phase is generally calculated to be the lowest- energy metastable phase for \BFO \cite{Dieguez/Gonzalez-Vazquez/Wojdel:2011,Dieguez/Iniguez:2011,Prosandeev/Wang/Ren:2012}), although competing phases can be lower energy depending on the choice of exchange-correlation functional \cite{Dieguez/Gonzalez-Vazquez/Wojdel:2011}.

We begin by searching for instabilities at the zone boundaries of the 20-atom $Pnma$ (GdFeO$_3$ structure) and $Pmc2_1$, and the 40-atom $Pbam$ unit cells, and construct the corresponding supercells by doubling them in one or two directions, resulting in unit cells of 40 or 80 atoms; larger unit cells will be presented in the next section. The $Pnma$ starting structure did not yield any new phases in this procedure, suggesting that it is likely a true metastable phase. We describe the structures generated from the $Pmc2_1$ and $Pbam$ starting cells next. When different structures share the same symmetry, we will label them with the symmetry followed by an index. 

\begin{center}\
\begin{figure}[h]\
\includegraphics[width=\columnwidth]{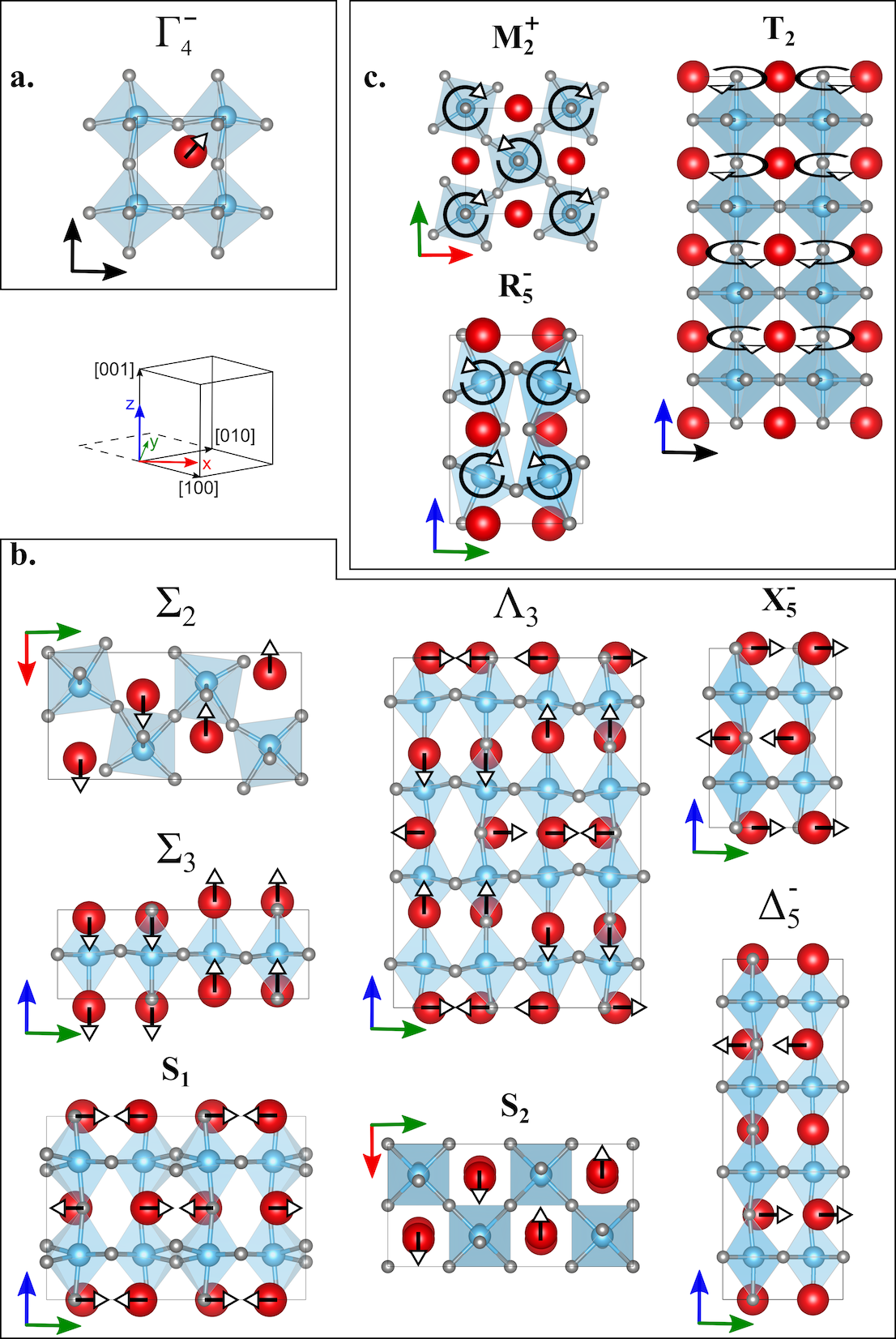}
\caption{\textbf{Main distortion modes - } (a) Polar mode dominated by off-centering of the Bi atom relative to the other atoms in the unit cell. Here we show the displacement along the $[011]_{pc}$ direction; displacements along  $[001]_{pc}$ (as in the $P4mm$ super- tetragonal phase) and along $[111]_{pc}$ (as in the $R3c$ phase) are also known. (b) Antipolar modes, in which pairs of Bi atoms displace with opposite direction and the same amplitude. (c) Rotational and tilting modes, in which the main distortion is the rotation or tilt of the octahedra. The inset shows the orientation of the axes related to the pseudo cubic directions and the dominant distortions are highlighted by arrows.}\
\label{fig_distortions}
\end{figure}\
\end{center}\

\subsubsection{$Pmc2_1$ as starting cell}\label{subsec_Pmc21}
We start from a 20-atom unit cell with $Pmc2_1$ symmetry and find a zone boundary instability along a long and a short lattice vector. Doubling its unit cell along the long lattice vector and subsequent relaxation of the atomic positions yields the 40-atom unit cell $Pc$ (1) structure (see Figure \ref{fig_structures} lower left), which is only 23 meV/f.u. above the ground state and 2 meV/f.u. above $Pnma$. It is a polar phase with an out-of-plane ($[001]_{pc}$) polarisation of around 54 $\mu$C/cm$^2$ and an in-plane ($[110]_{pc}$) polarisation of about 39 $\mu$C/cm$^2$. The polarisation is mainly due to the displacement of the eight Bi atoms, six of which shift along the $[1\bar{1}1]_{pc}$ direction and two along the $[\bar{1}11]_{pc}$ direction resulting in a larger out-of-plane than in-plane polarisation. 

By doubling the 20-atom  $Pmc2_1$ unit cell along $[\bar{1}10]_{pc}$ (short lattice vector), we identify a 40-atom unit cell polymorph with the same symmetry, $Pmc2_1$ (2),  which has a large polarisation along the $[110]$ pseudo-cubic direction. It is the highest energy phase that we identify here, with 76 meV/f.u. above the ground state and calculation of its phonon band structure indicates that this phase is unstable. Note that there is no zone boundary instability to motivate doubling the other short lattice vector. 

Further doubling the $Pmc2_1$ (2) unit cell along $[001]_{pc}$, we find a zone boundary instability. We subsequently relax the atomic positions and identify a metastable structure of $Cmc2_1$ symmetry with 80 atoms per unit cell (central right panel of Figure \ref{fig_structures}). There exists another larger unit cell with 80 atoms, $Pmc2_1$ (1), obtained by doubling the unit cell of the original $Pmc2_1$ simultaneously along $[001]_{pc}$ and $[110]_{pc}$ (lower right of Figure \ref{fig_structures}). $Pmc2_1$ (1) has a lower polarisation than $Pmc2_1$ (2) (16 $\mu$C/cm$^2$ instead of 76 $\mu$C/cm$^2$, due to a smaller amplitude of the $\Gamma_4^-$ mode), and lower energy (32 meV/f.u. instead of 76 meV/f.u.). The structural difference between $Cmc2_1$ and $Pmc2_1$ (1) lies in the presence of an extra antipolar displacement along x in the case of $Pmc2_1$ (1) that brings the central layer of Bi atoms (third layer of atoms along the z direction) to be almost aligned (see Figure \ref{fig_structures}). This difference is due to the absence of a $\Sigma_2$ mode (see Figure \ref{fig_distortions}b) in $Cmc2_1$. Moreover, the amplitude of the polar mode is larger in $Cmc2_1$ resulting in a larger polarisation than $Pmc2_1$ (1).

\subsubsection{$Pbam$ as starting cell}
Calculating the phonon band structure of the $Pbam$ (ground state for PbZrO$_3$ \cite{Sawaguchi/Maniwa/Hoshino:1951,Shirane/Sawaguchi/Takagi:1951}) phase, we first identify an unstable mode at the center of the Brillouin zone. Following this unstable mode, we obtain a non-polar phase, with $P2_1/c$ symmetry and an energy of 46 meV/f.u. above the ground state (upper left of Figure \ref{fig_structures}). In this phase $\Sigma_2$ and $\Sigma_3$ distortions are present, resulting in an up/down displacement of the Bi atoms both in-plane and out-of-plane and no net polarisation (see figure \ref{fig_distortions}).

Doubling the $Pbam$ unit cell along $[001]_{pc}$ and subsequently relaxing the atomic positions, we obtain two 80-atom structures. First, the $Pnma$ (1) phase, which was previously identified and shown experimentally to be antiferroelectric in Ref.~\onlinecite{Mundy/Heikes/Grosso:2018}.  And second, another variant of the same symmetry, $Pnma$ (2), which is metastable, but has much higher energy (70 meV/f.u.). 

It was shown that the $Pbam$ structure of PbZrO$_3$ can be described in termes of three main distortion modes from the ideal cubic perovskite structure, $\Sigma_2$, $S_2$ and $R_5^-$, which couple cooperatively to lower the energy \cite{Iniguez/Stengel/Prosandeev:2014}. $Pnma$ (1) and $Pnma$ (2) share the three modes inherited from their $Pbam$ parent structure, but also share two additional modes, $\Lambda_3$ and $T_2$. While the amplitudes of the three first modes are comparable in $Pnma$ (1) and $Pnma$ (2), those of $\Lambda_3$ and $T_2$ differ, representing respectively, 14$\%$ and 17$\%$ of the entire displacement from the cubic phase for $Pnma$ (1) and 2$\%$ and 12$\%$ for $Pnma$ (2).

\setlength{\tabcolsep}{10pt}
\renewcommand{\arraystretch}{1.2}
\begin{table*}[t]
  \centering
\resizebox{\textwidth}{!}{%
  \begin{tabular}{ l  r r r c c r r r}
  \toprule[1.5pt]
   &
  \multicolumn{3}{c}{\textbf{Lattice parameters}} &
  \multicolumn{2}{c}{\textbf{Energies}} &
  \multicolumn{3}{c}{\textbf{{Polarisation}}} \\ 

  \cmidrule[0.7pt](lr){2-4}
  \cmidrule[0.7pt](lr){5-6}
  \cmidrule[0.7pt](lr){7-9}

 \textbf{Phase}& a [\r{A}] & b [\r{A}] & c [\r{A}]  & Energy [meV/f.u.] & Band Gap [eV] & $P_a$ [$\mu$C/cm$^2$] & $P_b$ [$\mu$C/cm$^2$]& $P_c$ [$\mu$C/cm$^2$] \\
   \cmidrule[1.2pt](lr){1-9}
$Pmc2_1$ (2) & \emph{7.79} & 11.11 & 5.56 & 76 & 2.119 & 0 & 0 & 76.1 \\ 
$Pnma$ (2) & 5.57 & \emph{15.43} & 11.22 & 70 & 2.296 & - &- & - \\  
$P2_1/c$ & \emph{7.75} & 11.19 & 5.57 & 46 & 2.205 & - &- & - \\ 
$Cmc2_1$  & \emph{15.67} & 11.01 & 5.54 & 36 & 1.993 & 0 & 0 & 31.7 \\ 
$Pnma$ (1) & 5.53 & \emph{15.65} & 11.16 & 35 & 2.108 & - &- & - \\ 
$Pmc2_1$ (1) & \emph{15.60} & 10.93 & 5.58 & 32 & 1.959 & 0 & 0 & 16.0 \\ 
$Pc$ (1)& \emph{15.74} & 5.48 & 5.58 & 23 & 2.055 & 54.4 & 0 & 39.8 \\ 
$Pnma$ & 5.63 & \emph{7.73} & 5.40 &  21 & 1.825 & - &- & - \\ 
$R3c$ (hexag.)  & 5.54 & 5.54 & \emph{13.72} & 0 & 2.141 & 0 & 0 & \emph{100.4} \\ 

  \bottomrule[1.5pt]
  \end{tabular}
  }
  \caption{Computed lattice parameters, energy compared to the ground state, band gap energy and polarisation for newly identified structures within 100 meV of the $R3c$ ground state. In all cases the calculated angles between the lattice vectors are equal or very close to $90^\circ$. The lattice parameter value in italics indicates the axis along the cartesian $z$ axis; we refer to this as the  out-of-plane axis in the context of our later discussion of thin-film heterostructures. The polarisation is computed by summing over an average value of the Born effective charge for each ion multiplied by its displacement from its position in the high-symmetry reference structure. We show the values for the ground-state $R3c$ (in the hexagonal setting, in which the $c$ axis corresponds to the $[111]_{pc}$ direction) and for the high-temperature $Pnma$ structures for reference.}
  \label{tab_abc_E_P}
\end{table*}

\begin{table*}[t]
  \centering
\resizebox{\textwidth}{!}{%
  \begin{tabular}{ l  l l l c l l l }
  \toprule[1.5pt]
   & 
  \multicolumn{3}{c}{\textbf{{Main structural distortions}}} & 
  \multicolumn{4}{c}{\textbf{Octahedral tilts}} \\
  
  \cmidrule[0.7pt](lr){2-4}
  \cmidrule[0.7pt](lr){4-4}
  \cmidrule[0.7pt](lr){5-8}
  
 \textbf{Phase}& Polar modes & Anti-polar modes & Rotational/Tilting modes & Glazer & \multicolumn{3}{c}{Extended Glazer notation} \\
  
    \cmidrule[1.2pt](lr){1-8}
$Pmc2_1$ (2) & $\Gamma_4^-$ & $\Sigma_2$,$S_1$ & $R_5^-$,$M_2^+$& - &$a^{\bar{\alpha} \beta \bar{\gamma} \delta}$ & $a^{\bar{\alpha} \beta \bar{\gamma} \delta}$ & $c^{\bar{\alpha} \bar{\alpha}} $  \\
$Pnma$ (2) & - & $\Sigma_2$,$S_2$& $R_5^-$,$T_2$ &- & $a^{\bar{\alpha} \beta \bar{\gamma} \delta}$ & $a^{\bar{\alpha} \beta \bar{\gamma} \delta}$ & $c^{\bar{\alpha} \bar{\alpha} \beta \beta}$\\
$P2_1/c$ & - &  $\Sigma_2$,$\Sigma_3$,$S_2$ & $R_5^-$& - &$a^{\bar{\alpha} \beta \bar{\gamma} \delta}$ & $a^{\bar{\alpha} \beta \bar{\gamma} \delta}$ & $c^{\bar{\alpha} \beta} $  \\
$Cmc2_1$  & $\Gamma_4^-$ & $\Lambda_3$,$X_5^-$ & $R_5^-$,$M_2^+$ &- & $a^{\bar{\alpha} \beta \bar{\gamma} \delta}$ & $a^{\bar{\alpha} \beta \bar{\gamma} \delta}$ & $c^{\bar{\alpha} \bar{\beta} \bar{\beta} \bar{\alpha}}$\\
$Pnma$ (1) & - & $\Sigma_2$,$\Lambda_3$,$S_2$ & $R_5^-$,$T_2$ &- & $a^{\bar{\alpha} \beta \bar{\gamma} \delta}$ & $a^{\bar{\alpha} \beta \bar{\gamma} \delta}$ & $c^{\bar{\alpha} \bar{\alpha} \beta \beta}$\\
$Pmc2_1$ (1) & $\Gamma_4^-$ & $\Lambda_3$,$\Delta_5$,$X_5^-$,$\Sigma_2$ & $R_5^-$,$M_2^+$& - &$a^{\bar{\alpha} \beta \bar{\gamma} \delta}$ & $a^{\bar{\alpha} \beta \bar{\gamma} \delta}$ & $c^{\bar{\alpha} \bar{\beta} \bar{\beta} \bar{\alpha}}$  \\
$Pc$ (1) & $\Gamma_4^-$ & $\Delta_5$,$X_5^-$ & $R_5^-$,$M_2^+$,$T_2$ & - & $a^{\bar{\alpha}\beta}$ &$a^{\bar{\alpha}\beta}$ &$c^{\bar{\alpha}\beta \bar{\gamma} \bar{\delta}}$ \\
$Pnma$ & - & $X_5^-$ & $R_5^-$,$M_2^+$ & $a^-a^-c^+$ & $a^{\bar{\alpha} \alpha}$ & $a^{\bar{\alpha} \alpha}$ & $c^{\bar{\alpha} \bar{\alpha}}$\\
$R3c$ & $\Gamma_4^-$ & - & $R_5^-$& $a^-a^-a^-$& $a^{\bar{\alpha} \alpha }$ & $a^{\bar{\alpha} \alpha }$ & $a^{\bar{\alpha} \alpha }$ \\

  \bottomrule[1.5pt]
  \end{tabular}
  }
  \caption{Dominant structural distortions and octahedral tilt patterns for the newly identified low-energy phases. The distortion modes are reported with respect to the ideal cubic perovskite structure ($Pm\bar{3}m$) in all cases. We separate the modes into three categories: Polar, Anti-polar or Rotational/Tilting, and we only report the modes contributing to at least 4 $\%$ of the total distortion for each structure. An exception is made for the Polar modes, which are always reported to distinguish between polar and non-polar phases. The octahedral tilt patterns are given in terms of the traditional Glazer notation where possible, and the extended notation for all cases. We align the out-of-plane axis reported in Table~\ref{tab_abc_E_P} with the third axis in the Glazer notation.}
  \label{tab_modes}
\end{table*}

\subsection{Long-wavelength structures} \label{sec:nanometers_uc}
Having explored the phase space of 40- and 80-atom unit cells in the previous section, we next extend our systematic search to instabilities at lower q-points of the 20-atom $Pmc2_1$ unit cell, and correspondingly larger unit-cell structures.

\begin{figure*}[t]\
\includegraphics[width=\linewidth]{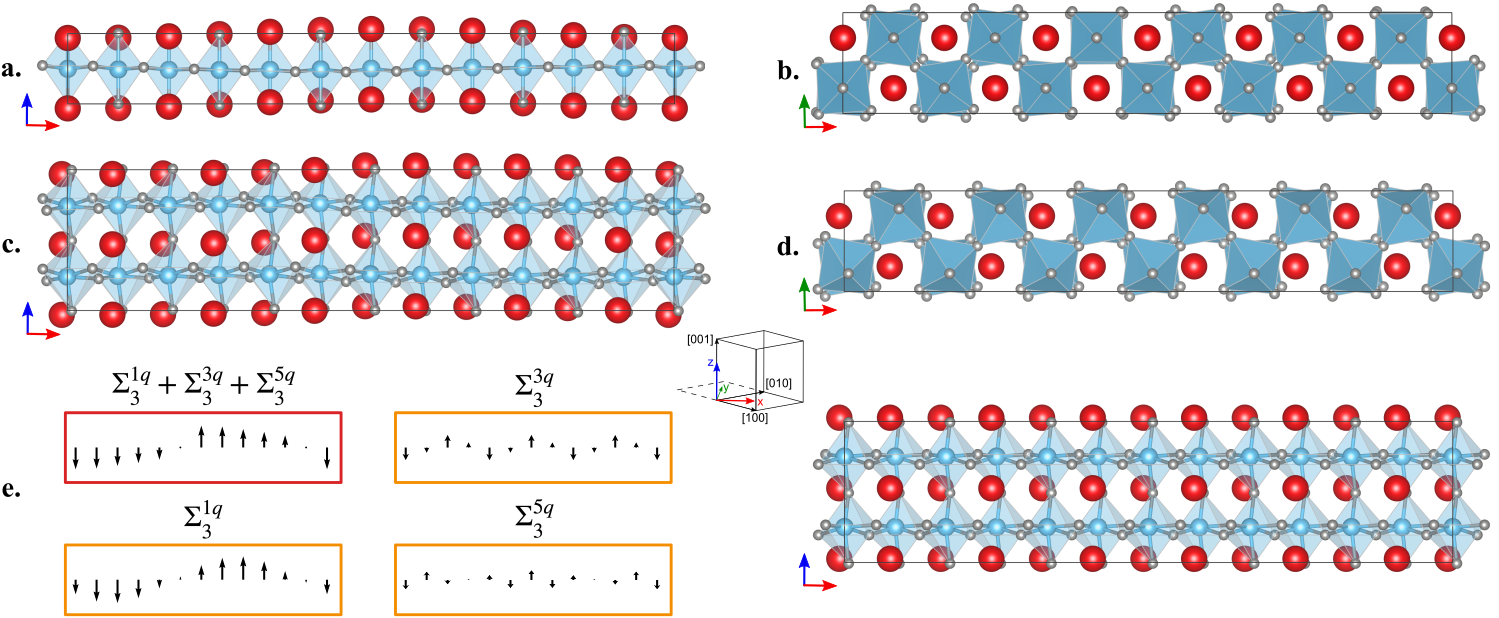}
\centering
\caption{\textbf{The long wavelength structural distortions and fully relaxed structure of the 33\text{ } \AA\text{ } long (120-atom)} unit cell - (a) Antipolar $\Sigma_3^q$ mode with sinusoidal displacement of the Bi atoms. (b) Rotational $S_4^q$ mode. (c) Fully relaxed structure with $Pc$ symmetry. (d) $R3c$ symmetry in the unit cell with same lattice vectors as (c) visualised from the top (upper panel) or from the side (lower panel). (e) Extra $\Sigma_3^{iq}$ modes with i = 1,3,5, individually presented (orange boxes) or combined together (red box). The structure in (c) can be decomposed into the modes presented in (a), (b) and (e) combined with a $\Gamma_4^-$ polar mode and an $R_5^-$ tilting mode. Each unit cell is defined with respect to a common origin, in order to help the comparisons and the orientation relative to the pseudo-cubic axes is indicated on the bottom left.}
\label{figure_longwavestruct}
\end{figure*}

We built a series of supercells, obtained by repeating the $Pmc2_1$ unit cell from 4 to 8 times along the $[110]_{pc}$ direction. Note that this is the direction for which no instability was found halway to the Brillouin zone boundary (see section ~\ref{subsec_Pmc21}). We also found no tripling imaginary zonecenter phonon modes in supercells built by the $Pmc2_1$ unit cell in that direction. We find, that an anti-polar distortion mode with $\Sigma_3$ symmetry is unstable in the starting supercells across the series, and is associated with a progressively smaller $q$-point in the Brillouin zone of the original $Pmc2_1$ phase as the supercell size is increased (see Figure \ref{figure_longwavestruct}a). In addition to this antipolar mode, we observe another wavevector-dependent mode, responsible for rotations of the octahedra, with $S_4$ symmetry (see Figure \ref{figure_longwavestruct}b). To emphasize the wavevector dependence of these modes we  label them $\Sigma_3^q$ and $S_4^q$. As previously, we freeze in the distortion eigenvectors corresponding to these and the other unstable modes and fully relax the structures. 

We find a series of structures with $Pc$ symmetry in unit cells with the long lattice vector ranging from 22.46\text{ \AA } to 44.78\text{ \AA } (corresponding to 80 and 160 atoms per unit cell, respectively). Each structure can be decomposed into four primary distortions: two modes already discussed in the previous section ($R_5^-$ and $\Gamma_4^-$) and the two new modes introduced above ($\Sigma_3^q$ and $S_4^q$). The $R_5^-$ mode is responsible for the tilts of the octahedra and is present in all the new structures identified in this work, including the $Pc$ series we discuss here. The $\Gamma_4^-$ mode displaces all the Bi atoms in the same direction relative to the other atoms, inducing polarisation in all polar structures. In contrast to the $R3c$ phase, the displacement in the new $Pc$ phases is not along [111]$_{pc}$ but along [110]$_{pc}$, resulting in purely in-plane polarisation. In Figures \ref{figure_longwavestruct} a-b, we show the new $\Sigma_3^q$ and $S_4^q$ modes for a 33\text{ \AA } long unit cell. Both modes have distortions with periodicity related to the size of their unit cell. $\Sigma_3^q$ is an antipolar mode, that moves the Bi atoms along the [001] direction. This mode has a wavevector of $q = [1/2\text{ }1/2\text{ }0] \cdot \frac{\pi}{N \cdot a}$, where $a$ is the lattice vector of the parent $Pmc2_1$ structure and $N$ the number of times it was repeated to generate the supercell. It is associated with an  
``$N$ up - $N$ down'' distortion pattern of the Bi atoms, with $N$ Bi atoms moving up and $N$ moving down, with the displacement amplitude modulated by the position along $[110]_{pc}$ by a sinusoid. 
The last mode, $S_4^q$, is a rotational mode that creates pairs of anticlockwise rotated octahedra along the [001] direction, with an amplitude of rotation modulated by a clipped sin wave along its long lattice vector (Figure \ref{figure_longwavestruct}b). 

The structure with 120 atoms is shown in Figure \ref{figure_longwavestruct}c and for comparison, the $R3c$ phase in the same unit cell is displayed in Figure \ref{figure_longwavestruct}d. Comparing carefully the anti-polar distortions in Figure \ref{figure_longwavestruct}a and Figure \ref{figure_longwavestruct}c, one can notice that the smooth wavy displacement due to the $\Sigma_3^q$ mode is not exactly conserved at the center of the unit cell in the latter. We observe that additionally to the $\Sigma_3^q$ mode with wavevector of $q = [1/2\text{ }1/2\text{ }0] \cdot \frac{\pi}{N \cdot a}$ as defined earlier, each structure across the series has additional $\Sigma_3^{iq}$ modes with $i = 3,5,7,...,N$. 
For instance in the structure with $q = [1/2\text{ }1/2\text{ }0] \cdot \frac{\pi}{6 \cdot a}$, two additional modes with $i= 3$ and $i=5$ exist. These additional modes have a decreasing amplitude as $i$ increases and they explain the deviation of the displacements of the Bi atoms from a smooth sinusoid as additional modes with shorter periods are superimposed (see Figure \ref{figure_longwavestruct}e).

\

\begin{figure*}[t]
\includegraphics[width=\linewidth]{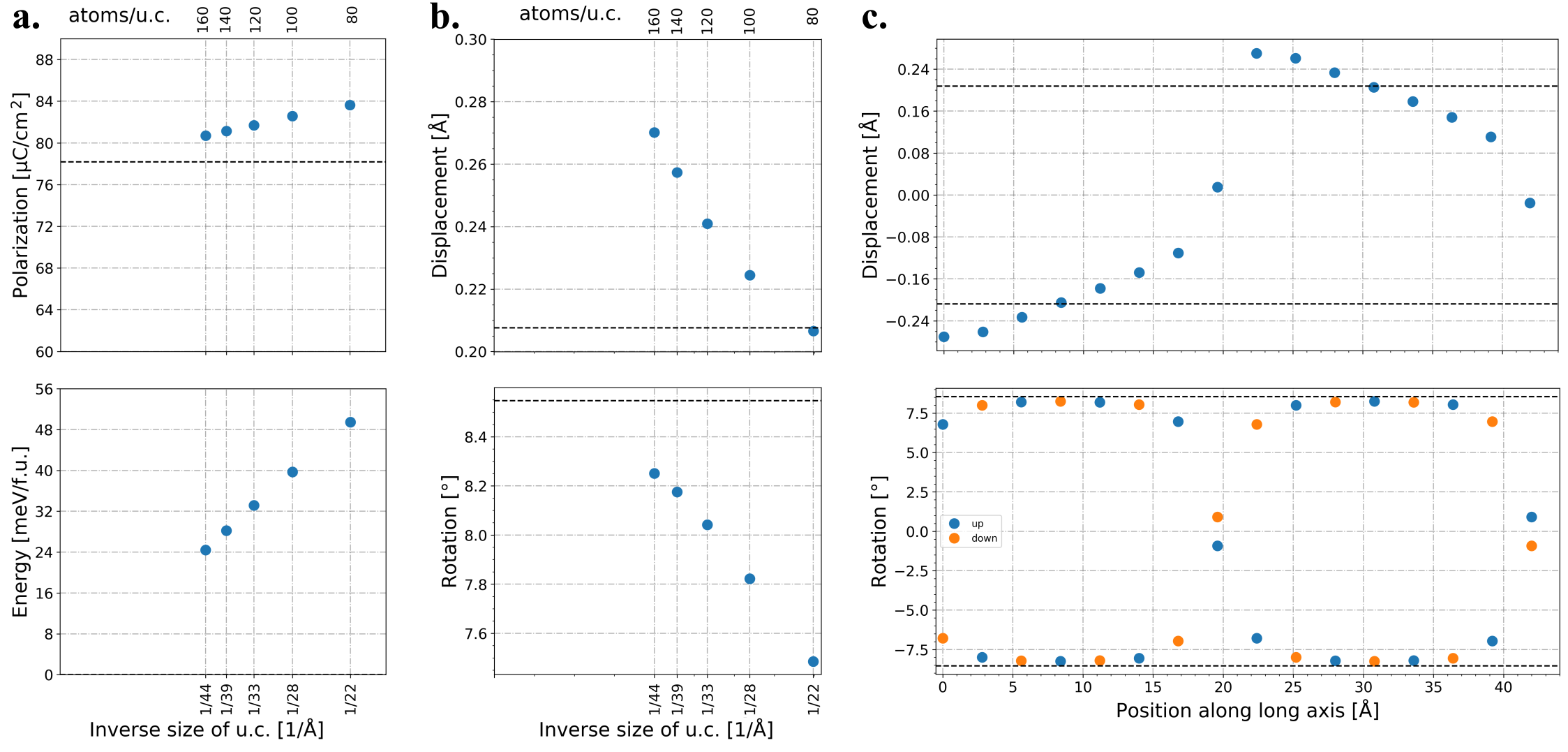}
\caption{\textbf{Energy, polarisation, Bi displacements and oxygen octahedral rotations across the $Pc$ series, and distortions within the 160-atom u.c.} (a) Polarisation (top) and energy (bottom) as a function of the inverse of the long lattice vector. (b) Maximum displacement of the Bi atoms along $[001]_{pc}$(top) and maximum rotation of the octahedra (bottom) as a function of the long lattice vector. (c) Layer-by-layer displacement of each Bi atom (top) and rotation of each octahedron (bottom) in the 160-atom unit cell. The orange and blue dots represent octahedra above (blue) or below (orange) the Bi atoms of the top panel. The dotted lines indicate the behavior of the $R3c$ in the same unit cell as the 160-atom unit cell.}
\label{fig_seriesPc}
\end{figure*}

In order to better understand the distortion trends in this series of structures, we plot in Figure \ref{fig_seriesPc}a the computed polarisation projected on the $[110]_{pc}$ direction and the energy relative to the energy of the $R3c$ phase constrained to the same lattice vectors as a function of the inverse of the long lattice vector. We observe that both values decrease and tend to the $R3c$ values as the unit cell size increases. This suggests that the series could converge to an analogue to the $R3c$ phase with in-plane polarisation. In Figure \ref{fig_seriesPc}b we plot the maximum displacement of the Bi atoms along $[001]_{pc}$ and the maximum rotation of the octahedra as a function of the inverse of the long lattice vector. We see that, as the unit cell size is increased, the maximum rotation value indeed tends to the value in the $R3c$ ground state. On the other hand, the maximum displacement of the Bi atoms along $[001]_{pc}$ (antipolar displacements) increases from the bulk $R3c$ value (polar displacements) as the cell size is increased. These trends are explained by the increase of both $\Sigma_3^q$ and $S_4^q$ modes when $q$ decreases. In a simple picture, as the octahedra have larger amplitudes of rotation, more space is left for the Bi atoms to move. These observations further indicate that the series in fact converges to a structure distinct from the $R3c$ ground state, although it is likely to be quite close in energy. Note that each half is also different than $R3c$.

We analyze the distortions further by plotting in Figure \ref{fig_seriesPc}c the displacements of each Bi atom along $[001]_{pc}$ as well as the rotation of each octahedron around the same direction, in the 160-atom unit cell. The values for the $R3c$ structure are shown with the dashed lines for comparison. In the case of $R3c$, all Bi atoms are displaced in the same direction (whether up or down) and the octahedra rotate clockwise in one (111) layer and anticlockwise in the next (111) layer (below or above the Bi atoms). In the $Pc$ phase, at the center of the unit cell, at around 20 \AA, one can see the crossover between up and down displacements of the Bi atoms, whereas at the center of the cell, there is almost no distortion: the Bi atom sits at the center of the cell (vertically) and the octahedra above and below are almost not rotated ($a^0$-like in Glazer notation). Going away from the center, the octahedra rotate in opposite directions to the left or to the right of the unit cell (with opposite rotations of the top and bottom layers) and the Bi atoms move up in the right and down in the left side of the unit cell. The bulk behavior is recovered, with displacements and rotations close to the $R3c$ values, midway between the center and the edges of the supercell, with opposite displacements / rotations in each half of the supercell.

As the unit cell size is increased, the amplitudes of the $\Sigma_3^q$ and $S_4^q$ modes grow. The rotations approach the ground-state $R3c$ value. The displacements become progressively larger, resulting in displacements larger than the bulk values. We attribute the lowering of the energy with increasing cell size to the smoother transition between consecutive octahedra, confirming previous studies \cite{Lubk/Gemming/Spaldin:2009,Dieguez/Aguado-Puente/Junquera:2013}, and anticipate that the energy would continue decreasing with increasing cell size.
Since the amplitude of the anti-polar displacements increases with the size of the unit cell, we expect large anti-polar distortions in the large unit cell limit. This could be interpreted as opposite domains of polarisation along $[001]_{pc}$, while preserving the in-plane components. Each domain keeps the $Pc$ symmetry but with displacements along $[001]_{pc}$ averaging to the $R3c$ value and resulting in a similar polarisation value. 

In summary, we have identified a series of structures that allow rotation of the polarisation away from the $[111]_{pc}$ direction and into the (110) plane at minimal energy cost. This behavior is relevant for \BFO thin-films and heterostructures in which the electrostatic boundary conditions might favor phases with no out-of-plane polarisation. We explore this scenario next.

\section{Stability and heterostructures} \label{sec:stability}

Finally, in this section we explore two possible routes to stabilizing the various phases identified above. The first is biaxial strain, which can be imposed by a substrate of different lattice constant through coherent heteroepitaxy. Since the different structures have different unit cell sizes and shapes, we expect that their relative stabilities will be modifiable by careful choice of in-plane lattice parameter lengths and orientations. The second is through exploiting the electrostatic boundary conditions at the interfaces in heterostructures or superlattices with non-polar III-III perovskite oxides. Since there is an electrostatic energy cost associated with a polar discontinuity, such an arrangement destabilizes phases with an out-of-plane polarisation component, including the ground-state $R3c$ phase in the usual [001] growth orientation \cite{Stengel/Iniguez:2015}. 

We continue the convention of Figures \ref{fig_structures} and \ref{fig_distortions} in which we oriented the usual $[001]_{pc}$ growth orientation along the cartesian z direction for all the structures. We therefore refer to the axis parallel to the z direction as the out-of-plane axis, and we apply in-plane strain it.

\begin{center}\
\begin{figure*}[th]\
\includegraphics[width=0.7\linewidth]{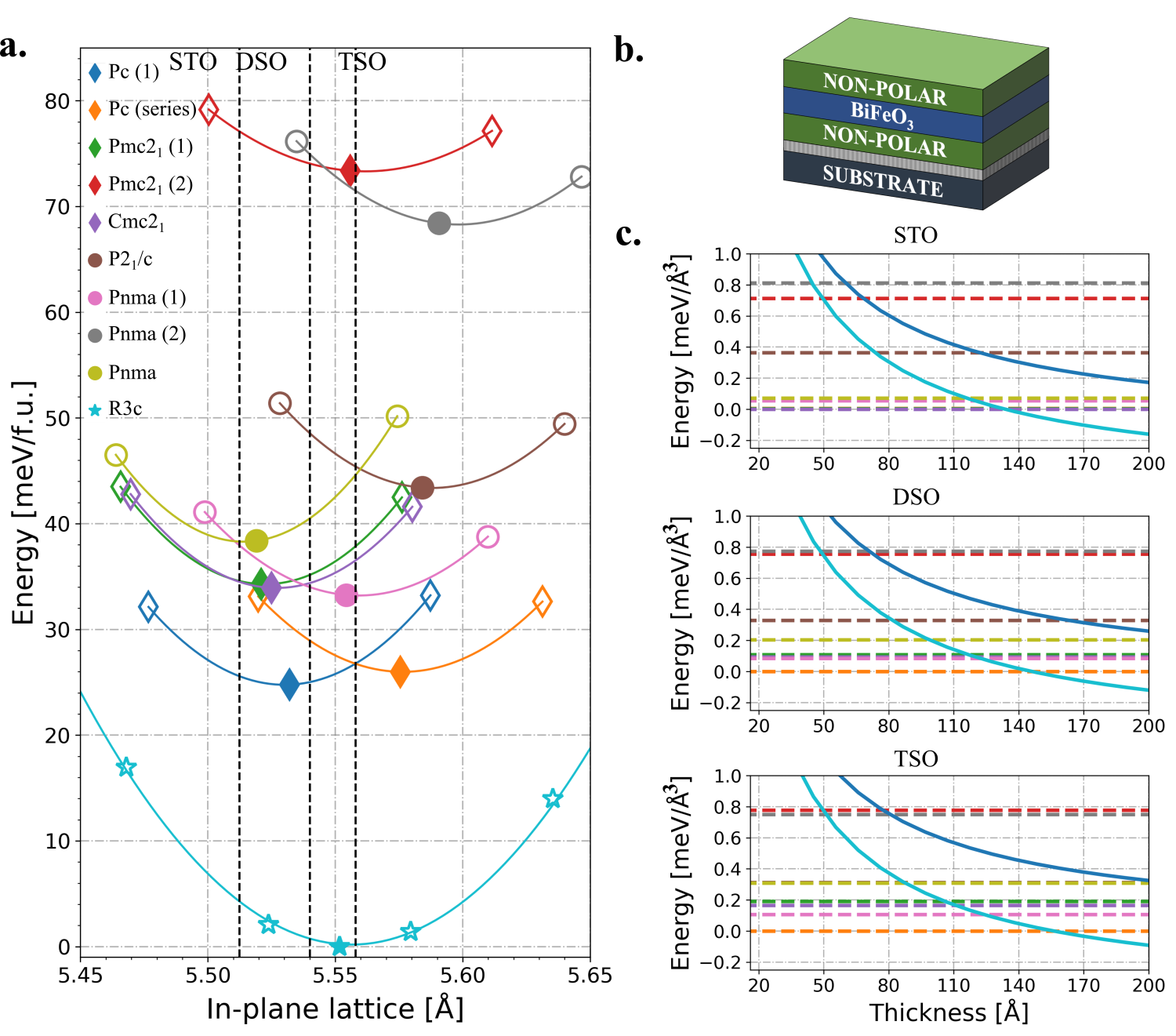}
\caption{\textbf{Strain and electrostatics - } (a) Energy as a function of the average in-plane lattice parameter. We consider each phase presented in Table \ref{tab_abc_E_P} and the $Pc$ phase presented in Figure \ref{fig_seriesPc}. For each case, we average the in-plane lattice constants and relax the out-of-plane vector as well as the positions of the ions. We indicate the resulting energies and lattice constants with full diamonds, full circles or full stars, for polar, non-polar and $R3c$ phases respectively. For the $R3c$ phase we used a 10-atom unit cell and kept the rhombohedral angle to be $60^\circ$ as described in Ref.  ~\onlinecite{Hatt/Spaldin/Ederer:2010}. We then apply a $1\%$ compressive and tensile strain to each structure and fit the data points with a parabola. The vertical dashed lines represent the lattice vectors calculated in this work of common substrates: SrTiO$_3$ (STO), DyScO$_3$ (DSO) and TbScO$_3$ (TSO). (b) Cartoon of the heterostructure considered for (c). We consider a superlattice of BFO surrounded by a non-polar III-III material (e.g. LaFeO$_3$) separated from the substrate (bottom) by a metallic electrode. (c) Energy as a function of the thickness of the BFO layer. We consider each of the phases presented in (a) and assume that they are surrounded by a III-III non-polar material with a strain imposed by the substrate (STO, DSO or TSO, respectively from top to bottom pannel). We report the energy density relative to the lowest non-polar phase. The colors are consistent with (a) and we indicate with dashed lines the phases that have zero component of the polarisation perpendicular to the interface. The volume and energy for each phase for a given substrate is extracted from (a).}
\label{figure_strain}
\end{figure*}\
\end{center}\

\subsection{Strain}

In Figure~\ref{figure_strain}a, we show the calculated energies for the phases of Table~\ref{tab_abc_E_P} as a function of their in-plane averaged lattice parameters. We make the choice to keep the angles between the lattice vectors fixed and average the pseudo-cubic in-plane lattice constants before applying the compressive or tensile strain, which increases the energy of the unstrained structures (full markers) compared to the values reported in Table~\ref{tab_abc_E_P}.

We find that, in the range of in-plane lattice parameters that we consider, none of the low-energy phases is stabilised by strain over the $R3c$ ground state. The energy difference between the low-energy competing phases and the ground state is sometimes substantially reduced on typical substrates however. For instance, the energy difference between $Pc$ (1) and $R3c$ is reduced from 28 meV/f.u. at the lattice constant of TbScO$_3$ (TSO) to 21 meV/f.u. at the lattice constant of SrTiO$_3$ (STO). The results shown in Figure ~\ref{figure_strain} show that, while moderate strain alone is insufficient to stabilize metastable phases, it makes a difference quantitatively and should not be neglected. Note that we only consider here strain values of less than $2\%$ (compressive or tensile) with respect to the lattice of the ground state , which are experimentally accessible for film thicknesses up to around 20 nm.

\subsection{Electrostatics and polar discontinuity}
According to Gauss's law, an electric charge causes a divergent electric displacement field with divergence equal to the charge density. Therefore, the bound charge at an interfacial polar discontinuity between a polar and a non-polar material will induce an electric field, with divergence proportional to the density of charge at the interface. This scenario is energetically unfavorable and is well known to be responsible for the critical thickness below which a ferroelectric material loses its net polarisation \cite{Junquera/Ghosez:2003}. In order to reduce its electrostatic energy, an unscreened ferroelectric material usually splits into domains of alternating up and down polarisation, resulting in a globally zero net polarisation. However, if a low-energy non-polar phase, or a phase with polarisation oriented in a plane parallel to the interface exist, a transition to a new phase could be more favorable than domain formation. This behavior was demonstrated for superlattices of \BFOo/(La,Bi)FeO$_3$, in which the phase labeled $Pnma$ (1) in this work was stabilized in the \BFO layers. 

Next, we explore whether the other low-energy phases identified in this work can be stabilized relative to the $R3c$ ground state with experimentally accessible electrostatic boundary conditions.

The bound charges induce a depolarising field, given by $E_d = \frac{-P}{\epsilon_0 \epsilon_r}$, in the opposite direction to the polarisation, where $\epsilon_0$ is the permittivity of the free space and $\epsilon_r$ the relative permittivity of the material in which the electric field is created. The depolarising field then couples with the polarisation, giving an electrostatic energy cost of $\frac{1}{2}\frac{P^2}{\epsilon_0 \epsilon_r}$ \cite{Marvan/Fousek:1999}. Note that this electrostatic energy cost is proportional to the interfacial area, whereas the internal energy, given by the energy difference between the two bulk phases at the appropriately strained in-plane lattice constant, is proportional to the volume. Therefore we expect a cross-over in their relative contributions with film thickness, with the electrostatic energy dominating in thin-films \cite{Stengel/Iniguez:2015}. 

We consider a heterostructure of \BFO surrounded by a non-polar material (e.g. LaFeO$_3$) separated from a substrate (STO, TSO or DSO) by a metallic electrode (see Figure \ref{figure_strain}b). In Figure \ref{figure_strain}c, we show the calculated energies of thin-films of the various phases of \BFO as a function of thickness, with the lattice constant fixed to that of the STO, TSO or DSO substrate, and the electrostatic boundary conditions set by imposing zero electric displacement field outside the \BFO layer. The calculated total energy consists of three contributions: the internal energy relative to the $R3c$ phase (including the effect of strain from the substrate), the electrostatic energy from the polar discontinuity at the interface, and a screening energy to decrease the amount of bound charge at the interface and reduce the electrostatic energy. As in Ref.~\onlinecite{Mundy/Heikes/Grosso:2018}, we assume that the system is single domain and take generation of electron-hole pairs across the band gap as the screening mechanism; this is likely to provide an upper bound value on the screening energy cost. 

Among the polar phases reported in Figure \ref{figure_strain}a, only $Pc$ (1) and $R3c$ have an out-of-plane component of the polarisation, 54 $\mu$C/cm$^2$ and 58 $\mu$C/cm$^2$ respectively, so their energy density decreases as a function of the thickness, while all other phases have a constant energy density, given by the internal energy. We find that $Pc$ (1) is unlikely to be stabilised on the considered substrates, as the smaller polarisation is not enough to compensate the strain energy compared to $R3c$. On STO, two phases with in-plane polarisation, $Pmc2_1$ (1) and $Cmc2_1$, are more stable than $R3c$ at low thicknesses. On the other hand, the largest structure from the series of $Pc$ phases is more stable than $R3c$ on DSO and TSO. We see that as expected, the critical thicknesses at which the phase with zero out-of-plane polarisation becomes stable is smaller when the difference in strain energy between the polar ground state and the lowest non-polar phase is smaller. 

We predict that \BFO on the different substrates considered would recover its bulk behavior only for thicknesses larger than 10 nm. As we mention, the electrostatic model that we chose is likely an upper bound estimation of the screening cost, which results in a likely overestimation of the electrostatic energy and critical thickness. Note also that due to the prohibitive cost of DFT calculations, we do not consider the competing mechanism of domain formation. This would be a fruitful direction for future second-principles or phase-field calculations. Nevertheless, our predictions give new routes to stabilizing new phases in \BFOo-based oxide heterostructures, and in general, as we showed that the control of the electrostatics at the interface can provide an extra control parameter, additional to the choice of substrate. 

\section{Summary and Conclusion} \label{sec:conclusion}
In summary, we used first-principles density functional theory to explore the low-energy phase space of \BFO systematically. We considered well-known polymorphs of \BFO and computed the phonon frequencies at high symmetry q-points to identify instabilities indicating structures of lower energy. Using this procedure, we showed that the previously discussed $Pbam$ and $Pmc2_1$ phases are in fact unstable. We first showed that several new metastable phases exist within an energy range of 50 meV/f.u. above the ground state in a unit cell of 80 atoms. These phases exhibit complex distortions and tilts that are beyond the scope of the traditional Glazer notation. We proposed an extension to the Glazer notation making it suitable for general tilt patterns. We then extended our study to larger unit-cell sizes by increasing the size of the original $Pmc2_1$ unit cell along one lattice vector up to a total cell size of 160 atoms. We discovered the existence of a series of non-centrosymmetric structures that are polar in the pseudocubic x-y plane, and antipolar in the perpendicular direction. We showed that the antipolar mode is modulated by the size of the unit cell and couples to a rotational mode that is similarly modulated. In the large supercell limit these phases are reminiscent of opposite domains of polarisation along the out-of-plane direction. Finally, we discussed the possibilities for stabilising these phases experimentally and showed that the lowest energy phases without out-of-plane polarisation could in principle be stabilised using a combination of strain and electrostatics at the interface between \BFO and a non-polar III-III perovskite. Future work could include an analysis of the role of other interfacial features such as matching of the tilt patterns of the oxygen octahedra in determining phase stability. 

Our approach of using phonon instabilities in a systematic way to explore the phase space of \BFOo, is of course broadly applicable to any material. It is limited, however, by the computational cost of the subsequent structural relaxations in the large unit cells. We hope, therefore, that our identification of many large-unit-cell low-energy phases will motivate further theoretical studies of the phase space of \BFO and related materials with modern tools such as machine learning or second-principles methods. These could reveal additional polymorphs with desirable properties such as antiferroelectricity. An extension to predict detailed magnetic properties would also be of interest, in particular to search for phases with a large net magnetic moment.

On the experimental front, we identified several new low-energy structures in \BFOo, as well as routes for stabilizing them, which we hope will be helpful in the design of new heterostructures with targeted properties. In particular, our study of the series of phases with $Pc$ symmetry showed that for large unit cells, reorientation of the polarisation is possible with minimal energy cost, and the distinction between phase and domains becomes blurred. The $Pc$ phases could therefore provide a starting point for engineering polar vortices \cite{Yadav/Nelson/Hsu:2016} in \BFOo-based oxide heterostructures.

\section*{Acknowledgments}
We acknowledge financial support from ETH Zürich and the Körber foundation. Computational resources were provided by ETH Zürich and the Swiss National Supercomputing Center (CSCS), Project ID No. s889. The visualisations of the structures were done with VESTA \cite{Momma/Izumi:2011} and we used Agate \footnote{{https://github.com/piti-diablotin/agate}{https://github.com/piti-diablotin/agate}} to compute the rotation angles for the octahedra. We thank Kane Shenton for helpful comments on the manuscript.

\bibliography{references}

\end{document}